# Imaging protein interactions in vivo with sub-cellular resolution

Valerica Raicu[1,2,*], Michael R. Stoneman[1], Russell Fung[1], Mike Melnichuk[1,†], David B. Jansma[1], Luca F. Pisterzi[3], Michael Fox[1], James W. Wells[3] and Dilano K. Saldin[1]

[1]*Department of Physics, University of Wisconsin, 1900 E Kenwood, Blvd., Milwaukee, WI 53211, USA*

[2]*Department of Biological Sciences, University of Wisconsin, Milwaukee, WI 53211, USA*

[3]*Leslie L. Dan Faculty of Pharmacy, University of Toronto, Ontario, Canada*

*e-mail: vraicu@uwm.edu

---

[†] Current address: University of Louisville, Louisville, KY 40292, USA

**Resonant Energy Transfer (RET) from an optically excited donor molecule (D) to a non-excited acceptor molecule (A) residing nearby is widely used to detect molecular interactions in living cells. Stoichiometric information, such as the number of proteins forming a complex, has been obtained so far for a handful of proteins, but only after exposing the sample sequentially to at least two different excitation wavelengths. During this lengthy process of measurement, the molecular makeup of a cellular region may change, and this has so far limited the applicability of RET to determination of cellular averages. Here we demonstrate a method for imaging protein complex distribution in living cells with sub-cellular spatial resolution, which relies on a spectrally-resolved two-photon microscope, a simple but competent theory, and a keen selection of fluorescent tags. This technology may eventually lead to tracking dynamics of macromolecular complex formation and dissociation with spatial resolution inside living cells.**

The theoretical and experimental underpinnings of Resonant Energy Transfer (RET) have been copiously explained in the literature, in the context of both fluorescence lifetime and intensity measurements [1-11]. While lifetime measurements on monomeric and associated proteins tagged with donor and acceptor fluorescent molecules may provide information regarding the distance between individual molecules in a complex, fluorescence intensities provide information on the interaction stoichiometry. In the latter case, an apparent RET efficiency ($E_{app}$) may be determined [8-11], which has been successfully used to extract the values of the fraction of associated vs. unassociated monomers and the number of monomers in a complex, as averages over several cells [5]. However, quantitative RET imaging faces significant difficulties, which have not been successfully addressed so far.

To better illustrate the challenges posed by quantitative intensity-based RET imaging and identify the requirements that the technology needs to meet, let us first write down the mathematical expressions for photon emission from donors in the presence of acceptors, ($F^{DA}$) and from the acceptors in the presence of donors ($F^{AD}$), respectively [8, 9, 12]:

$$F^{DA} = F^D(\lambda_{ex}) - F^D(RET), \tag{1a}$$

$$F^{AD} = F^A(\lambda_{ex}) + F^A(RET), \tag{1b}$$

where $F^D(\lambda_{ex})$ and $F^A(\lambda_{ex})$ are the fluorescence emissions from the donors and acceptors directly excited by laser light of wavelength $\lambda_{ex}$, $F^D(RET)$ is the loss of donor emission due to RET, and $F^A(RET)$ is the acceptor emission due to additional stimulation through RET (also known as *sensitized emission*).

Two similarly defined but quantitatively different *apparent RET efficiencies* are usually described in the literature on RET [5, 8, 9, 11]. One of these efficiencies quantifies the extent to which the donor is quenched by acceptors as a result of energy transfer, i.e.,

$$E_{app}^{Dq} \equiv \frac{F^D(RET)}{F^D(\lambda_{ex})}, \tag{2a}$$

where the superscript "$Dq$" stands for "donor quenching." Note that, while $F^{DA}(\lambda_{ex})$ may be easily determined experimentally using a spectral decomposition method (see below), determination of $F^D(\lambda_{ex})$ requires further consideration. This is because it is rarely possible to physically separate a natural oligomer into donor-tagged and acceptor-tagged components to determine the fluorescence intensity of the donor in the absence of energy transfer. However, it is often possible to inactivate (i.e., photobleach) the acceptor, by using a second excitation wavelength ($\lambda'_{ex} \neq \lambda_{ex}$) that leaves the donor unexcited and then measure $F^D(\lambda_{ex})$ by returning to $\lambda_{ex}$ [13, 14]. The second definition of RET efficiency quantifies the extent of sensitization of the acceptor by the excited donor, i.e.,

$$E_{app}^{Ase} \equiv \frac{F^A(RET)}{F^A(\lambda_{ex})} \qquad (2b)$$

where the superscript "*Ase*" stands for "acceptor sensitized emission." Experimental determination of $F^A(\lambda_{ex})$ in this equation is based again on exciting the acceptor at a wavelength $\lambda'_{ex} \neq \lambda_{ex}$ at which D is excited much less than A is, and inferring the intensity of acceptor emission upon excitation at $\lambda_{ex}$ in the absence of RET (i.e., $F^A(\lambda_{ex})$) from prior knowledge of the excitation spectrum of the acceptor.

Determination of RET efficiency from acceptor photobleaching suffers from the fact that significant photobleaching is also inflicted on donors. Depending on its severity, donor bleaching either precludes application of equation (2a) or introduces the need for sophisticated corrections [5]. In addition, once the acceptor is photobleached, its further use in kinetic studies of protein interactions is precluded. By contrast, determination of RET efficiency from sensitized emission (equation 2b) requires no irreversible bleaching of the fluorescent tags. However, since the direct excitation of acceptors is used as a reference for the efficiency defined in this way, $E_{app}^{Ase}$ depends on the instrument used and the excitation wavelength (since acceptor absorption depends on wavelength). Furthermore, equation (2b) is unusable in the experimentally desirable situation when the acceptor is not directly excited by incident light (so that acceptor emission is solely due to the interaction with donors), in which case $E_{app}^{Ase}$ tends to infinity. Another serious difficulty in using either of the classical methods for determination of RET efficiency with spatial resolution stems from the need to successively scan the sample at two excitation wavelengths. Because the time between such scans is long compared to the timescale of molecular diffusion in cells, the local composition of the sample may change from scan to scan, even though at whole cell level the composition may have remained the same.

## RESULTS AND DISCUSSION

IDENTIFICATION OF THE TECHNOLOGICAL REQUIREMENTS

At this point, the problem of quantitative RET imaging may be concisely stated: intensity measurements during a single scan of the sample provide only the two parameters on the left-hand-sides of equations (1), while computations of the apparent RET efficiencies [using equations (2)] for each pixel in the image require simultaneous knowledge of all the parameters on the right-hand-sides of equations (1). Careful evaluation of the physical problem allowed us to identify the requirements that a true image-pixel-level RET method needs to meet in order for this longstanding difficulty to be overcome, as discussed next.

    (*i*) In the practice of RET measurements, the short Stokes shift of most molecules faces the researcher with two mutually conflicting requirements: high overlap between the donor emission spectrum and acceptor excitation spectrum for increased RET efficiency, vs. good discrimination between the emission of D and that of A to allow for accurate computation of the RET efficiency. The conflict arises specifically from the fact that close overlap between the D emission and A excitation brings about significant overlap between their emission spectra (due to the relatively short Stokes shift of most fluorescent molecules). One of the more popular choices has been to relax the condition of high spectral overlap so that the requirement of discriminating detection may be met using optical filters. This choice has the advantage of simplicity, but leads to low RET efficiency and, hence, to reduced accuracy. A relatively small but growing community of researchers in the

RET microscopy field relies instead on spectral decomposition of measured spectra [5, 15-17], while keeping RET efficiency high, for increased accuracy of the measurements. With this method, the left-hand-sides of equations (1) may be determined precisely using spectral decomposition of spectra containing both donor and acceptor signals (see Methods section). The method provides two quantities, $k^{DA}$, $k^{AD}$, which are equal to the maximum emission intensity for D in the presence of A, and for A in the presence of D, respectively. These may be used to determine the total number of photons emitted per unit of time by integrating the fluorescence spectra, $I^{DA}(\lambda_{em}) = k^{DA} i^D(\lambda_{em})$ and $I^{AD}(\lambda_{em}) = k^{AD} i^A(\lambda_{em})$, over all emission wavelengths, i.e.:

$$F^{DA} = k^{DA} \int_{\lambda_{em}} i^D(\lambda_{em}) d\lambda_{em} = k^{DA} w^D, \qquad (3a)$$

$$F^{AD} = k^{AD} \int_{\lambda_{em}} i^A(\lambda_{em}) d\lambda_{em} = k^{AD} w^A, \qquad (3b)$$

where $i^A(\lambda_{em})$ and $i^D(\lambda_{em})$ are the emission intensities of the pure A and pure D spectra normalized to their maximum values, and $w^A$ and $w^D$ are the integrals of the elementary spectra of the acceptor and donor, respectively. Note that the total number of quanta emitted may be determined from equations (3) only if entire fluorescence spectra are available and not just two different wavelengths, as proposed in some spectrally-resolved RET studies.

(*ii*) It is also necessary that all parts of the emission spectrum originate from the same molecules or molecular complexes in the sample. Since biological samples evolve in time, the molecular composition of a region of the cell may change before intensity at all emission wavelengths is recorded, if serial detection is used. This does not constitute a significant limitation if the information is analyzed in terms of averages over the whole cell. However, in RET microscopy with sub-cellular resolution, parallel acquisition of fluorescence intensities at all wavelengths of the spectrum is mandatory.

(***iii***) Another important requirement of the new technology is to selectively excite the donors with negligible excitation of acceptors [i.e., $F^A(\lambda_{ex}) \cong 0$]. This allows one to reduce to just three the total number of unknowns in equations (1). In the present work, this is accomplished simply by relying on the selectivity of two-photon excitation, which is high compared to the single-photon excitation [18].

Reduction of the whole problem of RET-based imaging to determination of just three unknown parameters in equations (1), by using the above conditions, would in itself constitute an important technological feat. However, a problem still remains: determination for each image pixel of the three parameters on the right-hand-side of equations (1) from the two measured parameters on the left-hand side is an obvious mathematical impossibility. Fortunately, a third equation may be obtained, which relates $F^D(RET)$ to $F^A(RET)$, as described next.

THE RELATION BETWEEN $F^D(RET)$ AND $F^A(RET)$ AND THE RET EFFICIENCY

Let us begin derivation of this equation by noticing that a fraction $Q^D$ of the total number of excitations, $N^{RET}$, transferred through RET to the acceptor would have been detected as photons emitted by the donor if RET were absent, according to the relation:

$$N^{RET} Q^D = F^D(RET). \tag{4a}$$

Once transferred, however, a fraction $Q^A$ of those excitations will be emitted as photons by the acceptor; mathematically, that situation may be expressed by the relation:

$$N^{RET} Q^A = F^A(RET). \tag{4b}$$

Note that the quantities $Q^A$ and $Q^D$ in equations (4) are in fact the quantum yields of the acceptors and donors, respectively. Substituting $N^{RET}$ from equation (4b) into (4a) we obtain the third necessary equation,

$$F^D(RET) = \frac{Q^D}{Q^A} F^A(RET), \quad (1c)$$

which, together with equation (1b) and the design principle (*iii*) [i.e., $F^A(\lambda_{ex}) = 0$], gives:

$$F^D(RET) = \frac{Q^D}{Q^A} F^{AD}. \quad (5)$$

This result is very general, and its derivation did not invoke any assumption regarding the nature of the interaction between D and A (such as oligomeric vs. stochastic interactions) or the size or type of oligomers (homo-oligomer vs. hetero-oligomer). Finally, combination of equations (1a), (2a), (3a), (3b), and (5) leads to the following expression for the apparent RET efficiency:

$$E_{app}^{Dq} \equiv \frac{F^D(RET)}{F^D(\lambda_{ex})} = \frac{1}{1 + \frac{Q^A}{Q^D} \frac{k^{DA}}{k^{AD}} \frac{w^D}{w^A}} \quad (6)$$

Equation (6) may be used to determine the RET efficiency directly from the measured fluorescence spectra of each image pixel and without recourse to acceptor photobleaching or multiple excitation wavelengths.

THE DESIGN OF A TWO-PHOTON MICROSCOPE WITH SPECTRAL RESOLUTION

The optical setup of our microscope [19] that meets all the above requirements is presented schematically in Fig. 1. Its excitation scheme is similar to most two-photon microscopes [20, 21]. A solid-state CW laser (Verdi™, Coherent Inc., CA) emitting at 532 nm pump a modelocked

Ti:Sapphire laser (KM Labs, CO), which generates femtosecond light pulses with wavelength ranging from about 700 nm to 860 nm, and FWHM ranging from ~30 nm to over 120 nm. The near-IR beam is focused to a diffraction-limited spot on the sample by an infinity-corrected plan apochromat objective (Nikon Instruments Inc. NY, 100× magnification, NA = 1.40, oil immersion) and raster-scanned across the sample using galvanometric scanners (Nutfield Technology Inc., NH). Fluorescence from the sample, collected using the same objective, is passed through a dichroic mirror and a short-pass filter (T = $10^{-8}$ for $\lambda \geq 670$ nm, Chroma Technology), to remove residual back-reflected excitation light, and then projected through a transmission grating onto a cooled electron-multiplying charge-coupled device (EM-CCD, Ixon DV 887, Andor Technology) with single-photon sensitivity. The scanners and the CCD camera are controlled by a computer using software written in house (in $C^{++}$). All lenses used are antireflection-coated achromatic doublets.

The novel feature of this microscope consists in the way in which spectral resolution is achieved. The projection of the emission spectrum onto the CCD camera lies in one direction (e.g., the *y*-direction), while the laser beam is scanned along the direction perpendicular to the spectrum (*x*-direction). The emission emanating from the sample along one line scan forms a rectangle on the CCD chip, and one such 2-D image is captured for each line scanned on the sample and saved on a computer. After performing line scans for different *y*-values of the sample the resultant images are reconstructed as described in the Supplementary Online Material to give the final spectrally-resolved fluorescence images. The current resolution is between 2 and 310 wavelength windows covering wavelengths from 400 to 650 nm.

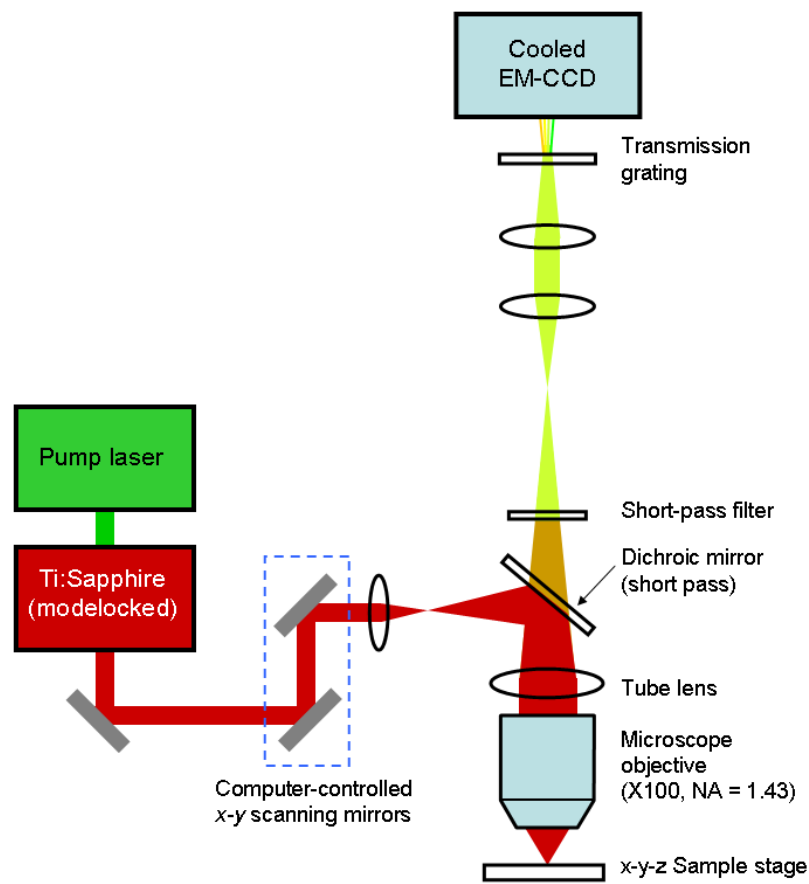

**Figure 1 Schematic representation of the two-photon microscope with spectral resolution** (see text for details).

When using fluorescent molecules with very short Stokes shifts, it is possible to achieve selective excitation using a pulse shaper [22] for coherent control [23-25]. In this work, we relied on the known discriminating ability of two-photon processes [18], and on a selection of the fluorescent molecules [17]. We used the variant $GFP_2$ [17] of the Green Fluorescent Protein (GFP), whose single-photon excitation spectrum is blue-shifted ($\lambda_{ex}^{max} \cong 400\,\text{nm}$) compared to the more common EGFP variant and to the yellow fluorescent protein [5, 7] which is used as an acceptor herein.

QUALIFICATION OF THE METHOD

To test the spectral-resolution capability of our microscope as well as its ability for selective excitation, we first used the instrument to image yeast cells (*S. cerevisiae*) genetically engineered to express fluorescent proteins in their cytoplasm (see the Methods section). One type of cells expressed GFP$_2$ in their cytoplasm while the other sample contained cells expressing YFP. Using light from the femtosecond laser with an average power of ~10 mW at the entrance of the microscope objective, and excitation wavelength centered around 800 nm (FWHM ~ 30 nm), we acquired fluorescence images of the GFP$_2$-expressing cells, which provided the emission spectrum of the GFP$_2$. Typical images (see Fig. 2) showed an emission spectrum for GFP$_2$ very similar to the one obtained with single-photon excitation of this fluorescent protein [17].

Next, we subjected the YFP-expressing cells to excitation light with same spectrum but twice the average power, which should give an increase by a factor of four in fluorescence [21]. By selecting for cells showing high levels of YFP expression (as judged from their overall emission intensity), we were able to obtain fluorescence images (Fig. 2) with average intensities (over several tens of cells) a factor of about five lower than those obtained from cells expressing GFP$_2$.

Normalized fluorescence spectra of YFP and GFP$_2$, averaged over all pixels from several cells, were used for spectral un-mixing of the composite spectra from cells expressing the fusion protein GFP$_2$-YFP, as described in the Methods section. When applied to the samples in Fig. 2, the pixel-level deconvolution method gave two-dimensional maps of the quantities $k^{DA}$ and $k^{AD}$ defined above (see the first two columns of Fig. 3). As expected, samples expressing only donor (GFP$_2$) or only acceptor (YFP) showed signals of only one type, i.e., $k^{DA}$ or $k^{AD}$, respectively. This confirms the accuracy of the spectral deconvolution method. By contrast, the samples expressing the GFP$_2$-YFP construct showed both $k^{DA}$ and $k^{AD}$ signals. The direct excitation of the YFP by excitation light accounted for less than 5 % of the YFP emission (see above).

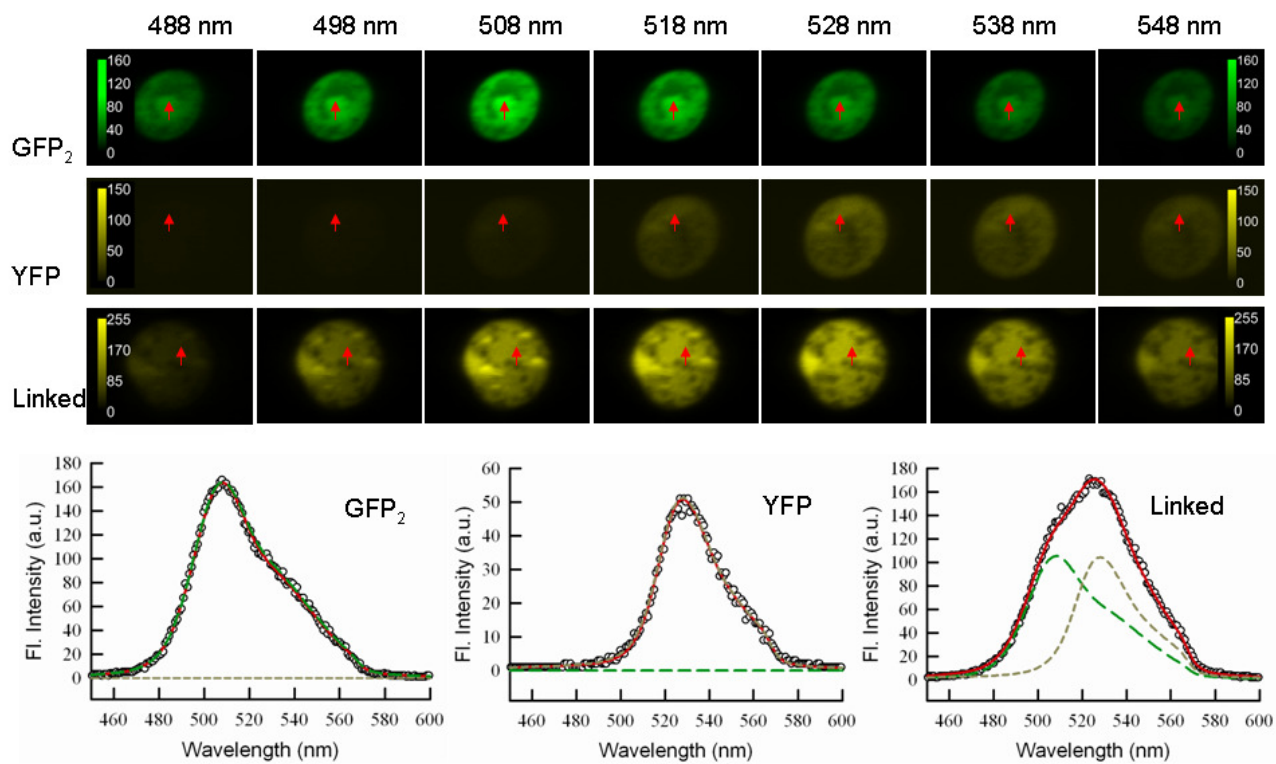

**Figure 2 Typical results obtained with the TPM in Fig. 1 from yeast cells expressing in their cytoplasm GFP$_2$, YFP, and GFP$_2$ linked to YFP. Top three rows**, Images obtained at seven selected emission wavelengths (out of a total of 310 wavelengths) upon excitation with laser pulses having ~30-nm bandwidth, centered about 800 nm. Note that the cell expressing YFP only was not typical (i.e., it was selected for high YFP expression), and that the average excitation power was two times higher than that used for the other cells, since YFP is barely excited at 800 nm (see text for details). **Bottom row,** Fluorescence spectra corresponding to the pixels indicated by arrows in images at the top. Circles represent the measured intensities, solid lines are best-fit spectra given by equation (8), long-dash lines correspond to deconvoluted donor-only spectra and short-dash lines correspond to acceptor-only spectra (see Methods sections).

The $k^{DA}$ and $k^{AD}$ maps thus determined were subsequently used to compute the maps of apparent RET efficiencies for all three types of sample, using equation (6), the known ratio of the quantum yields ($Q^D/Q^A = 0.55/0.61 = 0.90$) [5, 17], and the ratio of the integrals of the normalized spectra ($w^D/w^A = 1.21$) determined from our measurements. All the computations were done using Mathcad (Math Soft, Inc), and required that the original 16-bit images be converted into 8-bit grey-level images (i.e., values ranging from 0 to 255). To avoid obtaining pseudo-RET efficiencies for pixels that only present background noise, only the grey levels that exceeded a value of 20 were considered for $E_{app}$ calculations. The $E_{app}$ map for all three types of samples is shown in the last column of Fig. 3. As expected, only the cell expressing linked proteins presented nonvanishing $E_{app}$ values. It is also noteworthy that the $k^{DA}$ map in Fig. 3 presented some four brighter regions that corresponded to almost completely dark regions in the $k^{AD}$ map and, therefore in the $E_{app}$ map. These regions are likely locations in the cell (endoplasmic reticulum, ER) where proteins are synthesized. Because the maturation time of the YFP is generally longer than the maturation time of green variants, the YFP molecules (acceptor) are probably not mature enough at those locations to emit light and to participate in RET with $GFP_2$.

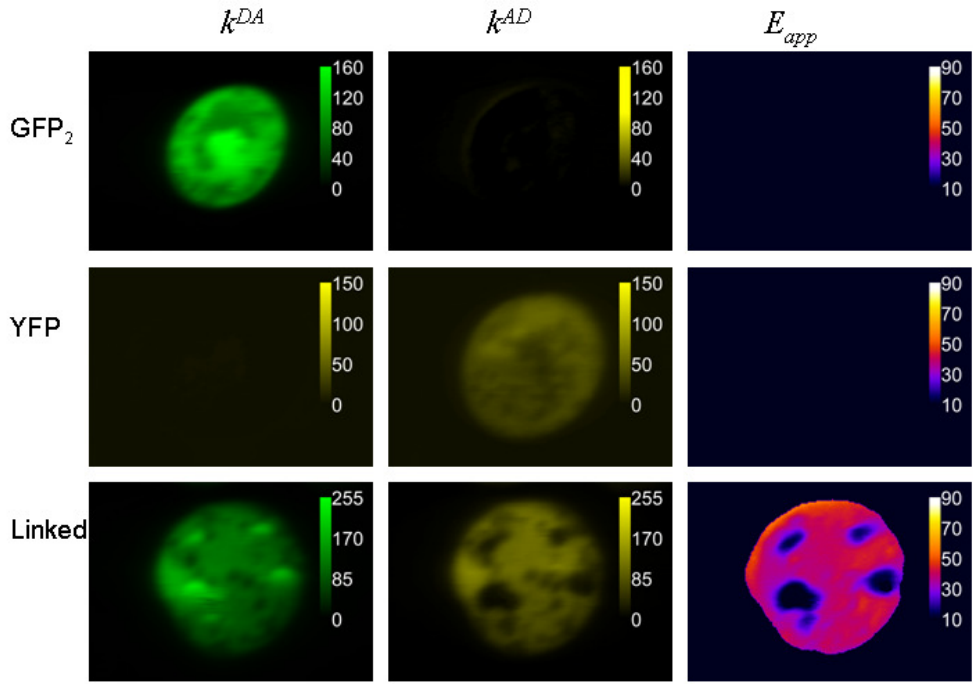

**Figure 3 Results of the analysis of the data in Fig. 2 with the theory described in the text. First two columns,** 2-D maps of donor ($k^{DA}$) and acceptor ($k^{AD}$) signals obtained from the images in Fig. 2 using a spectral deconvolution procedure described in the Methods section. The intensity is in arbitrary units. **Third column,** 2-D maps of apparent RET efficiencies ($E_{app}$) computed from images to the left using equation (6) and expressed in %. Intensities are assigned false colors, according to their values and the scales shown as insets in each image.

To investigate whether the RET efficiencies thus obtained are physically acceptable, we also determined the average (over multiple measurements) RET efficiency for the GFP$_2$-YFP construct from fluorescence lifetime imaging (FLIM) [26], as described in the Methods section. The FLIM-based method gave a true (or pair-wise) RET efficiency of (44 ± 2) %. Since, by design, our molecules "form" heterodimers, the apparent RET efficiency and the true RET efficiency should take on similar values [12]. The average $E_{app}$ for the whole cell shown in Fig. 3 was ~35%,; values of $E_{app}$ fell abruptly in some dark areas of the cells (presumably, at the ER, due to the lack of mature, optically active acceptors) and in some small areas, where $E_{app}$ reached ~50%. The larger values may either reflect uncertainties associated with this method or, more likely, slight contributions from stochastic RET [3, 27, 28] due to high concentration of the proteins in some regions of the cytoplasm. Given the presence of completely dark regions, the average value of the $E_{app}$ for the cell compared well to that determined from FLIM. We performed similar determinations of RET efficiency from over 40 cells expressing the GFP$_2$-YFP construct, and then averaged the results from all cells. A mean of (39 ± 10) % was obtained for the apparent RET efficiency which, once again, was in good agreement with the value determined from FLIM.

## CONCLUSION

When higher order oligomers (such at trimers, tetramers, etc) are present in the cell, the apparent efficiency should be higher than the true efficiency. With the use of an adequate theory for RET efficiency in mixtures of complex and free monomers [12, 29], the two different $E_{app}$ values will provide the size of the oligomer and the proportion of associated and unassociated proteins for every image pixel. Finally, with further increase in the scanning speed, which is achievable using some available scanning protocols [30], this method will allow one to monitor protein complex dynamics in living cells.

## METHODS

SAMPLE PREPARATION

Variants of the green fluorescent protein (GFP) linked with two amino acids have been used in this work as an RET probe. GFP$_2$ has been used as a donor of energy, which has excitation and emission spectra similar to the wild-type GFP with the difference that the excitation peak at 398 nm is more pronounced for $GFP_2$, which is caused by the amino-acid substitution F64L [17]. The acceptor molecule was a variant of the GFP with red-shifted emission spectrum, called yellow fluorescent protein. It contained the following amino-acid substitutions (compared to the wild type): S65G S72A T203Y. The DNA encoding $GFP_2$ was generated by site-directed mutagensis, introducing into a wild-type sequence of GFP from *Aequorea victoria* (a gift from J. Greenblatt) the nucleotide change which resulted in an F64L amino-acid substitution in the protein. YFP was a gift from K. J. Blumer. The nucleotide sequence TCCGGA (which encodes the amino acids Serine and Glycine), was attached by PCR to the C-terminus of the coding sequence of $GFP_2$ and the N-terminus of the coding sequence of YFP. These two PCR fragments were combined with vector DNA in living yeast cells by homologous recombination [31]. The resulting plasmid, expressed two GFP variants linked by the amino acids SerGly. The cells containing the plasmids of interest were frozen for subsequent use. For determination of the fluorescence spectra of $GFP_2$ and YFP, these two proteins have been singly expressed in the cells. This has been done by using plasmids that only contained $GFP_2$ or YFP separately.

Yeast strains incorporating one of the above three plasmids were grown on synthetic-complete solid medium lacking uracil in order to select for plasmids of interest. Galactose was used as a carbon source since the linked fluorescent molecules were expressed from the GAL1 promoter. Yeast colonies were suspended in water and deposited on solid medium as a drop that was allowed to form a streak by tilting the plate. The plate was then incubated overnight at 30 °C. A cell sample consisting of ~3 μl of 0.02 M KCl solution and cells gently scraped from a plate was placed on a microscope slide, covered with a glass coverslip and then placed on a *x-y-z* microscope stage for fluorescence measurements. The KCl solution and the residual glucose helped maintain a neutral pH, which ensured constant quantum yield of the fluorescent molecules during measurements [5].

SPECTRAL CALIBRATION OF THE MICROSCOPE

A simple calibration was implemented to assign the appropriate wavelength to each reconstructed fluorescence image. A full set of line scans were performed upon a standard sample of which the fluorescence emission spectrum was known. The scanning parameter $\Delta y$ was set such that the fluorescence spectrum of the sample shifted by exactly one pixel upon one increment of the x-y scanning mirrors in the y direction (spectral dimension). The standard sample consisted of a 100-µM solution of fluorescein sodium salt (Uranine, Fisher Scientific, IL) dissolved in deionized water. Utilizing a function of the open source code software Image J (http://rsb.info.nih.gov/ij/), the reconstructed images, indexed by integer $i$, could be opened and stacked in ascending order. By organizing the reconstructed images in this manner, the background corrected normalized fluorescence intensity of each reconstructed image was determined as a function of $i$. The relationship between camera pixel position in the spectral dimension and the wavelength of the emitted photons was approximately linear; therefore the relationship between the reconstructed images and corresponding wavelength was also linear and calculated as follows:

$$\lambda_i = m(i - i_{\max}) + \lambda_{\max} \qquad (7)$$

where the value of $m$ is represented by:

$$m = \frac{(\lambda_{max} - \lambda_{1/2})}{(i_{max} - i_{1/2})}.$$

In the above formalism, the symbol $\lambda_i$ represents the wavelength of the $i^{th}$ reconstructed image. The values of $\lambda_{max}$ and $\lambda_{1/2}$ extracted from the emission spectrum of the fluorescein correspond to the wavelengths at which the fluorescence intensity of the fluorescein sample is a maximum and half of maximum, respectively. The image numbers $i_{max}$ and $i_{1/2}$ correspond to the reconstructed images possessing the maximum fluorescence intensity and one half of the maximum, respectively, and were determined as stated above.

SPECTRAL DECONVOLUTION

In experiments with cells expressing *only A or only D molecules*, fluorescence intensities were measured at 310 wavelengths, $\lambda_{em}$ (uniformly spread between 400 and 650 nm), to obtain fluorescence spectra for both species (see Fig. 2). All the intensities in each spectrum were divided by the maximum intensity in that spectrum, and then averaged over all pixels in a cell and over several cells to obtain normalized fluorescence spectra, $s^A(\lambda_{em})$ and $s^D(\lambda_{em})$. The fluorescence intensities of the sample containing *both A and D* were measured at the same wavelengths to obtain the composite fluorescence spectrum $S(\lambda_{em})$. The measured spectrum is related to normalized $s^A(\lambda_{em})$ and $s^D(\lambda_{em})$ spectra through the following relation:

$$S(\lambda_{em}) = k^{DA} s^D(\lambda_{em}) + k^{AD} s^A(\lambda_{em}), \tag{8}$$

where $k^{DA}$ and $k^{AD}$ are the individual signals from D in the presence of A and from A in the presence of D. $k^{DA}$ and $k^{AD}$ were determined using a least-squares procedure [5].

DETERMINATION OF RET EFFICIENCY FROM FLUORESCENCE LIFETIME IMAGING

Time-correlated single photon counting (TCSPC) was carried out on budding yeast in a solution of 200 mM KCl with a Zeiss Axiovert 100 laser-scanning confocal microscope equipped with Hamamatsu R3809U-52 detectors controlled by a Becker-Hickl DCC-100 control module. Cells were imaged under a coverslip (Fisher Scientific) using a 100x oil immersion lens, and the excitation source was a Chameleon multiphoton laser system (Coherent) tuned to 900 nm. Cells were scanned for a period of 480 s on power setting 8. Emission from $GFP_2$ was selected using a FF02-460/80-25 (Semrock) centred at 460 nm with a bandwith of 80 nm.

The acquired FLIM images were analysed on a pixel-by-pixel basis using SPC Image (Becker-Hickl, Germany). The fluorescence decay ($\tau$) from each pixel was fit with a single exponential decay, $f(t) = a_0 + a_1^{-t/\tau}$, where $a$ is the amplitude and $t$ corresponds to time. On average, the pixels were binned five times, to increase the number of photons per curve. This yielded a distribution of the lifetimes present in the sample. These distributions then were fit with a Gaussian function, which took the functional form of

$$f(x) = Ae^{-\frac{(x-\tau)^2}{2\sigma^2}} \quad (9)$$

The parameters $A$ and $\sigma$ correspond to the amplitude and standard deviation of the function. The fit was optimized in terms of $A$, $\tau$ and $\sigma$. The optimization of parameters was carried out as described previously [32]. The centre of the Gaussian, was taken as the mean lifetime and subsequently used in calculations of the true FRET efficiency:

$$E = 1 - \frac{\tau_{DA}}{\tau_D} \quad (10)$$

where $E$ is the true RET efficiency and $\tau_{DA}$ and $\tau_D$ are the lifetimes of $GFP_2$ in the presence or absence of YFP, respectively [9].

For the cells expressing donor alone, the average lifetime (weighted by (Gaussian's area/ S.E.$^2$) was (2.35 ± 0.05) ns (N = 49 cells), while the average lifetime for the cells expressing the GFP$_2$-YFP construct was (1.31 ± 0.02) ns (N = 47 cells). These values gave $E = 0.44 ± 0.02$.

**Acknowledgements**

This work was supported by a grant from the Wisconsin Institute for Biomedical and Health Technology (Grant No. W620), and seed funds from the UWM Research Growth Initiative (Grant No. X014). We thank Devin Gillman for useful suggestions regarding the computer routines for instrument control, and Michael J. Woodside of the Imaging Facility at Sickkids Hospital for his assistance with the lifetime imaging.

**Author contributions**

V.R. designed the TPM, built the optical setup, participated in some of the experiments, developed the RET theory, the algorithms for image reconstruction, and the computer routine for image analysis, and wrote the bulk of the paper. M.S. and M.F. carried out the experiments using the TPM with spectral resolution and analyzed the data. R.F. wrote computer codes for controlling the galvanometric scanners and the EM-CCD camera, and for image acquisition and reconstruction. M.M. interfaced the scanners with the computer and built an earlier version of the optical setup. D.B.J. designed and made the genetic constructs. L.P. and J.D.W. determined the RET efficiency using FLIM measurements. D.K.S. participated in theoretical modeling. All authors contributed to manuscript writing and editing.

**Supplementary online material on image reconstruction**

By using a standard sample that fluoresces uniformly across the sample (e.g., fluorescein), the line-scanning procedure can be calibrated so that from line-scanning at one y-value to the next on the sample (let us say, from $y = y_0$ to $y = y_0 + \Delta y$) the fluorescence spectrum moves by exactly one pixel along the spectral dimension on the detector, as described in the Methods section. This calibration procedure also allows identification of the relation between row numbers and wavelengths on each of the images as shown in supplementary figure. The reconstruction procedure is illustrated in supplementary Fig. To obtain the fluorescence emission image for a particular wavelength, let us say, $\lambda_5$, one finds the row number on the first image (i.e. the first y) that corresponds to this wavelength (row 5 in this example), then the next row of the next image (i.e. row 6 on the second image) would, by virtue of the calibration procedure, correspond to the same wavelength but for the next y, and so on. Stacking all the image rows that correspond to the wavelength of interest, an image of the fluorescence emission of the sample at the wavelength of interest is obtained. This procedure is then repeated for other wavelengths.

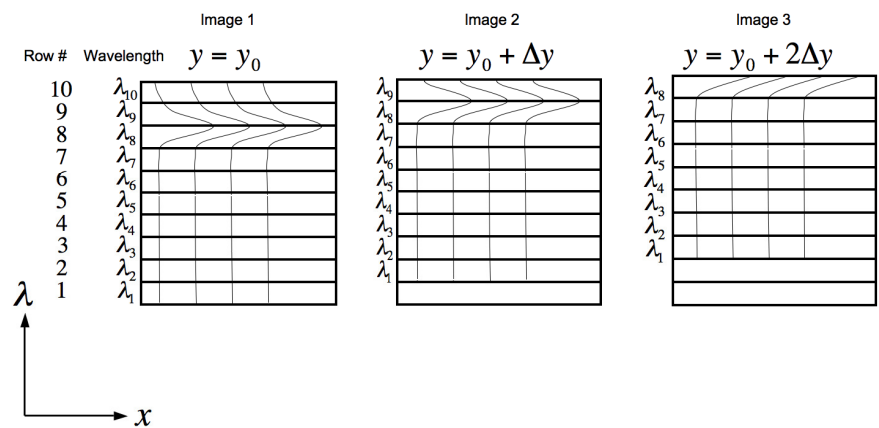
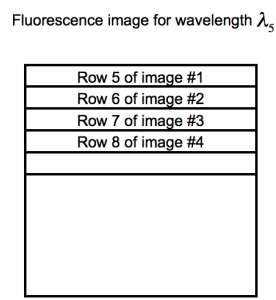

**Supplementary figure** (a) Illustration of the calibration procedure. (b) Reconstruction of the image from individual scans.